\documentclass[preprint]{elsart}
\usepackage{graphicx}
\usepackage{textcomp}
\usepackage{amsfonts}
\usepackage{amssymb}
\journal{arXiv}
\begin{document}
\begin{frontmatter}
\title{Evolutionary Optimization of State Selective Field Ionization for Quantum Computing}
\author{M.~L.~Jones},
\author{B.~Sanguinetti},
\author{H.~O.~Majeed},
\author{B.~T.~H.~Varcoe}
\address{School of Physics and Astronomy, University of Leeds, Leeds, LS2 9JT}
\maketitle
\begin{abstract}
State selective field ionization detection techniques in physics require a specific progression through a complicated atomic state space to optimize state selectivity and overall efficiency. For large principle quantum number $n$, the theoretical models become computationally intractable and any results are often rendered irrelevant by small deviations from ideal experimental conditions, for example external electromagnetic fields. Several different proposals for quantum information processing rely heavily upon the quality of these detectors. In this paper, we show a proof of principle that it is possible to optimize experimental field profiles in situ by running a genetic algorithm to control aspects of the experiment itself. A simple experiment produced novel results that are consistent with analyses of existing results.
\end{abstract}
\end{frontmatter}
\section{Introduction}
Many problems in experimental physics involve optimizing a set of parameters in a large state space. Theoretical models exist for many problems of interest, but they are typically only valid with the application of a series of approximations, many of which do not necessarily hold true in a real laboratory situation. Complications arise, for example, when a stray electromagnetic field is present that perturbs the system of interest. Such fields are often impossible to remove completely from an experiment, and so it is frequently found that the measured parameters differ from the predictions. 

The field of quantum computing has expanded rapidly in recent years to encompass many different fields with its promise of exponential or polynomial speedup in a number of important computations, such as factoring \cite{Shor1994} and searching \cite{Grover1997}. To build a useful quantum computer, it is necessary to gain exquisite control over individual quanta (e.g.~atoms, electrons or photons) in a system. The advantage of quantum computing derives directly from the fact that extremely large state spaces can be manipulated in non-trivial ways to perform certain calculations very much more efficiently than their classical counterparts. Evolutionary techniques have been successfully applied to some problems in quantum computing, in particular in the design of circuits to perform quantum logic operations \cite{Barnum2000}. There exist several proposed schemes for implementing quantum computing, for example ion traps \cite{Cirac1995}, nuclear magnetic resonance (NMR) \cite{Gershenfeld1997} and cavity quantum electrodynamics \cite{Blythe2006}.
\section{Quantum Computing in Cavity Quantum Electrodynamics}
The field of cavity quantum electrodynamics (cavity QED) has proved itself to be one of the leading candidates for quantum computing, since it offers many advantages, including the ability to couple light and matter in a controlled way. Current state of the art implementations of cavity QED experiments involve passing a sequence of carefully prepared atoms through a superconducting cavity that houses a microwave field containing small numbers of photons. For a number of technical reasons, so-called \emph{Rydberg} states \cite{book:Gallagher2005} are used, which means that a valence electron has been promoted via laser excitation to very high principal quantum number $n$. This leaves the electron loosely bound to the nucleus allowing the atom to be easily ionized in an electric field. The nature of the interaction is such that only two different Rydberg states participate, so the ability to distinguish between them after the cavity is sufficient to allow us to perform measurements for quantum computation. As in many other cutting edge atomic physics experiments, the atoms are typically alkali metals, in our case rubidium.

\section{State Selective Field Ionization}
\begin{figure}[!ht]
\begin{center}
\includegraphics[width=0.7\textwidth]{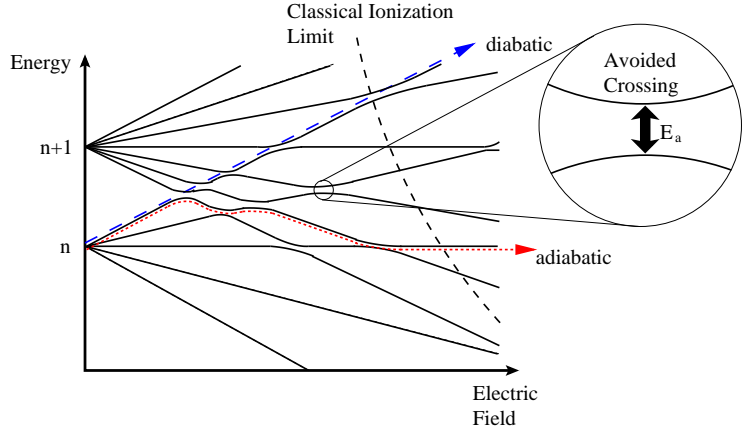}
\caption{Simplified schematic of a typical Stark map. As the electric field is increased, different energy levels are shifted by different amounts. Adiabatic (dotted arrow) and diabatic (dashed arrow) paths are shown. The inset shows the structure of an avoided crossing, with energy gap $E_a$.}\label{fig:stark}
\end{center}
\end{figure}

Applying an electric field to an atom causes its energy levels to be Stark shifted to different energies. The operating principle of field ionization detectors is that it is possible to apply a varying electric field to `walk' an energy level to the ionization limit, at which point the electron escapes and is collected in an electron multiplier, producing a detector `click'. Different starting states reach ionization at different electric fields. Fig.~\ref{fig:stark} shows a simplified schematic of the effect of an electric field on the energy levels of the atom.
Due to the interaction of the valence electron with the core of an alkali atom, features known as avoided crossings are exhibited in the Stark maps. The size of these avoided crossings defines the energy required to jump from one level to the next. If the electric field is ramped slowly with respect to this energy gap, the path will tend to follow an adiabatic path (dotted arrow in fig.~\ref{fig:stark}) and ionize exactly at the classical ionization limit. If the field is changed quickly with respect to this energy gap, it is likely that the avoided crossings are traversed diabatically (dashed arrow in fig.~\ref{fig:stark}), and the atom is ionized at a point at or above the classical ionization limit. Intermediate slew rates, however, result in a random combination of adiabatic and diabatic traversals, leading to the ionization field becoming much less well defined, which significantly reduces the state discrimination of the detector. The states of interest in our experiment, around $n=60$, are in this intermediate regime and therefore effective detection has been a persistent problem.

The Rydberg states have an average lifetime of the order of 0.1--1ms, so it is also necessary to complete the field ramp before the atom decays from the Rydberg state. The detector must also fit inside the the physical space available inside the experimental vacuum chamber. This places constraints on both the maximum and minimum timescales for the different measurements.

Traditional designs of field ramp for single state selective detectors have tended to involve an approximately linear ramp, with more complicated combinations of linear ramps of differing slew rate to distinguish between different states. 

\section{Technical details of the experiment}
\begin{figure}[!ht]
\begin{center}
\includegraphics[width=0.7\textwidth]{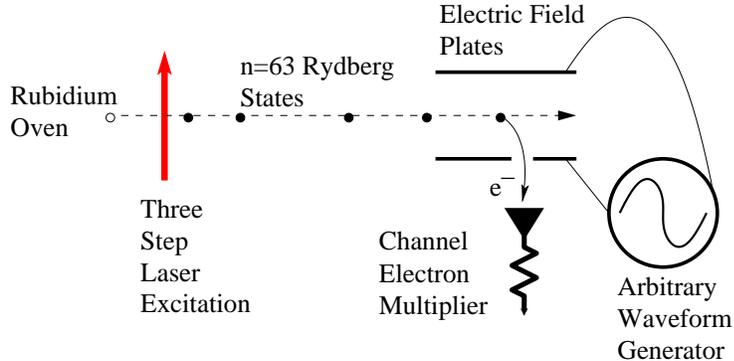}
\end{center}\caption{The experimental setup that was optimized by the GA. The laser excitation is a sequence of three infrared lasers at 780nm, 776nm and 1257nm. A time varying voltage is applied to the field plates and ionized electrons are collected at the channel electron muliplier (CEM).}\label{fig:experiment}
\end{figure}

The experimental arrangement controlled by the GA is shown in fig.~\ref{fig:experiment}.
A beam of rubidium atoms are excited to high-lying Rydberg states via a three step laser excitation process. These atoms pass through an ionization region where the ionization is driven by the electric field between a pair of parallel plates. The liberated electrons are collected in a channel electron multiplier (CEM) which records a TTL pulse for each detection event. The atomic beam and detectors are contained within a vacuum chamber at a pressure of around $10^{-8}$mbar. The laser light is coupled into the chamber via optical fibers.

This arrangement is a subset of the quantum optical \emph{micromaser} experiment \cite{Walther2006} that probes fundamental quantum mechanical effects, such as the controlled creation of arbitrary numbers of photons \cite{Varcoe00} and entanglement \cite{Englert1998}. The frequencies of the first two laser steps are locked to the relevant atomic transitions inside a vapor cell on the optical table. It is not possible, however, to lock the third step in this way, and for this experiment this laser was locked with an optical frequency synthesizer (Menlo FC1500). 

\subsection{Temporally varying vs.~spatially varying profiles}
For this particular experiment, we chose to use a temporally varying field (sometimes referred to as pulsed field ionization \cite{book:Gallagher2005}) since it is technically easier to achieve, requiring only a single electrical feedthrough into the vacuum system. To ensure that we are only detecting atoms that are in the correct portion of the field ramp, it is necessary to gate the electron multiplier's output. Clearly this duty cycle puts a limit upon the maximum count rates that will be measured. 

\subsection{The Problem to be solved}
The first experiment that has been performed, as a proof of principle, involved simply a beam of atoms prepared in a particular Rydberg state (in this case a state with $n=63$) impinging upon an externally controlled field ionization region with a single electron multiplier.

Broadly speaking, there are two different scenarios that a detector will be required to cope with. First of all, many cavity QED experiments require a fixed and very precise atomic velocity, requiring a detector to be optimized very tightly to that velocity, perhaps at the expense of other velocities \cite{Jones2009}. The majority of experiments will, however, require the ability to detect atoms moving at a wide range of velocities. For example, a typical experiment will consist of a sequence of atoms travelling at identical velocities, followed by one or more travelling at a different velocity as a probe of the quantum state inside the cavity \cite{Varcoe00}.

To optimize for the first type of experiment, we have two schemes for velocity control at our disposal. First of all, we can align the laser excitation at an angle to the atomic beam to ensure we only excite atoms with the correct Doppler shift (and hence correct velocity). The second scheme is a time of flight style system where the laser excitation is pulsed and the CEM gate occurs at a fixed time afterwards.
\begin{figure}
 \begin{center}
 \includegraphics[width=0.8\textwidth]{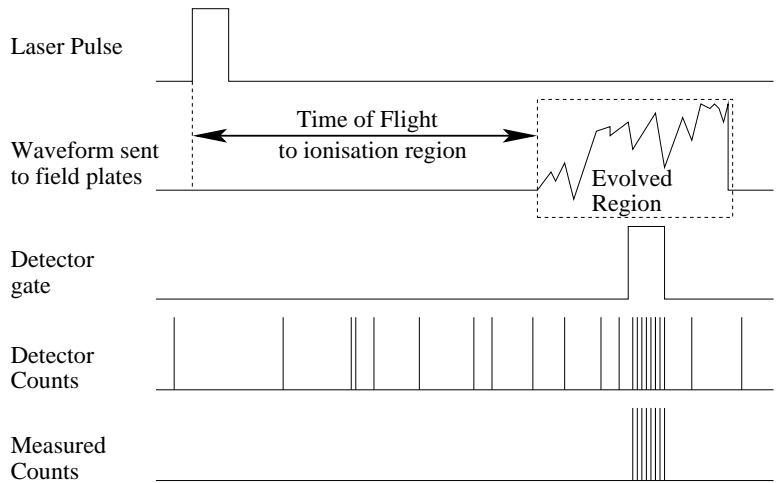}
\end{center}\caption{Pulsed excitation scheme. The laser excitation is pulsed on for a short period, exciting a small group of atoms moving at a wide range of different velocities. Only those atoms that are ionized during the gate pulse are recorded, hence selecting only a very narrow velocity class.}\label{fig:pulses}
\end{figure}
This type of scheme is illustrated in fig.~\ref{fig:pulses}.

For the second type of experiment, we require the evolved solution to be robust to varying atomic velocities. This is achieved by simply preparing the full Maxwell-Boltzmann distribution of atoms and optimizing the detector simultaneously to all velocity classes present in the atomic beam.
\begin{figure}
 \begin{center}
 \includegraphics[width=0.8\textwidth]{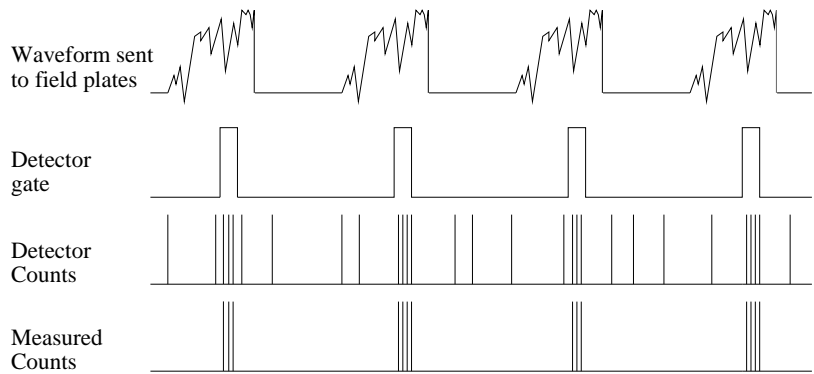}
\end{center}\caption{Continuous excitation scheme. The laser excitation is applied continuously so that atoms of all velocities are present in the detector during the gate pulses. This scheme allows much faster repetition rates and hence offers better counting statistics.}\label{fig:continuous}
\end{figure}
To do this we apply a continuous perpendicular excitation to the atomic beam and gate the electron multiplier synchronously with the evolved waveform, as shown in fig.~\ref{fig:continuous}. This allows us to gate the detector many more times a second and evaluate the fitness of a given individual with correspondingly higher confidence.

A possible combination of the two schemes would be to create a catalogue of profiles optimized for different velocities and actively switch between them for the velocity of the next atom.

A spatially varying profile has a number of advantages over a simple temporally varying profile. The main advantage is that there is no need to gate the electron multiplier's output signal, since every atom experiences the same field, regardless of the time at which it arrives. This immediately gives us a gain over the limited duty cycle necessary in a temporally varying system.

There is also a subtle difference in that the spatially varying system will produce a slew rate that is a function of the atomic velocity, unlike the temporally varying system, where a given instantaneous electric field value is experienced at all positions inside the detector. 

\section{The Genetic Algorithm}
The genetic algorithm used in this case is based upon the \emph{Microbial GA} \cite{Harvey2001}. In summary, this scheme of GA utilizes a pairwise tournament selection, the parametric uniform crossover is implemented by overwriting elements of the tournament loser with the corresponding  elements from the tournament winner, with probability $P_c=0.5$. A standard creep style mutation is employed, with the mutation size $X$ being drawn from a normal distribution centered on $\langle X\rangle=0$V with standard deviation $\sigma_X=5$V.

The initial population consisted of 50 individuals of 20 evolvable vertices that were initialized with random values uniformly distributed over the range of -45V to 25V.
\subsection{Fitness Function}
A key element of any GA is the choice of fitness function that drives the selection procedure. In this simple example of the experiment, the fitness function is chosen to be the absolute count rate, since a higher count rate indicates a higher detection efficiency. Future designs will simultaneously optimize for two different atomic states, and hence will require a correspondingly more complex fitness function.
A key advantage of a tournament style of GA is that the pairwise selection helps to negate any effects of medium to long term variations in the experiment. For example, slight laser drift may cause the flux of atoms in the 63P state to fluctuate, which would make a generational scheme unreliable, as the best individuals won't necessarily achieve the best fitness. Since these types of variations typically occur on a timescale longer than the time to evaluate just two individuals (a few seconds), we can be confident that the tournament winner actually is the better individual.
\subsection{Genotype}
The genotype that was evolved was a set of vertices that defined a waveform which was fed to an arbitrary waveform generator (\emph{Tektronix AFG3022}). The voltage output of this device was then amplified and fed to the parallel electrodes inside the vacuum chamber (fig.~\ref{fig:experiment}). The time varying voltage applied to these plates defines the electric field profile in the detection region.
\section{Results and Analysis}
\begin{figure}
 \begin{center}
 \includegraphics[width=0.9\textwidth]{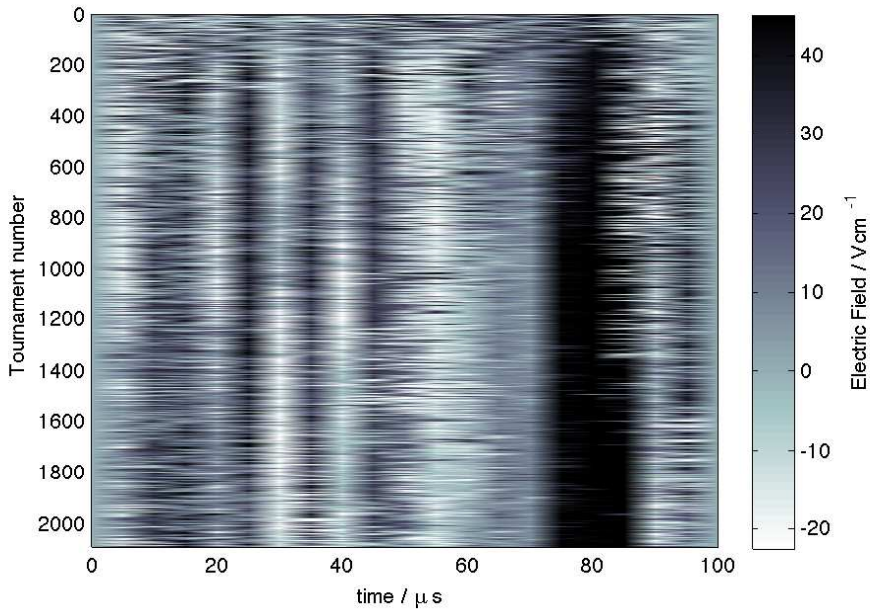}
\end{center}\caption{The winner of each tournament during the evolution. The initial random population quickly develops a series of sharp oscillations followed by a plateau where the detector gate is active, between 75--85{\textmu}s.}\label{fig:results}
\end{figure}

The GA was allowed to run for approximately 2000 tournaments, with each evaluated individual being recorded along with its fitness. The results are plotted in fig.~\ref{fig:results}, showing that the expected fast field ramp at the time of the detector gate (75--85{\textmu}s) appears very early in the evolution, as might be expected. More unexpectedly, further ramps appeared in the field profile before the  detector gate.
The presence of a single negative dip before a large positive pulse has been noted before in a similar experimental system \cite{Tada2002}, the explanation for which is that the atoms are manipulated into a state that preferentially ionizes exactly at the classical limit. The presence of multiple dips before the main pulse in our experiment suggests that the system is seeking a state which is more easily ionized. The more complicated structure most probably occurs due to the different value of $n$ we use, placing us in the intermediate slew rate regime as opposed to the much higher $n$ used in reference \cite{Tada2002}, which is firmly in the purely diabatic regime.
\section{Future Directions}
The first extension is to incorporate a second detector and optimize for state selectivity. This would involve an extended fitness function that provides selection pressure to increase the efficiency of each detector individually and also for their ability to distinguish the correct state. The system must be cooled to cryogenic temperatures (as it is in the micromaser experiment) in order to suppress thermally induced intermixing of the Rydberg states that would introduce noise into the fitness evaluation.

In typical experiments, the experiment is run continuously, so a pulsed detection scheme is unsuitable. In this case, a spatially varying field is required. To accomplish this, we are in the process of building a new detector where the single pair of electrodes are replaced by an array of strip electrodes. The voltages sent to each of these electrodes will then be the genotype to be evolved.


\begin{thebibliography}{10}

\bibitem{Shor1994}
P.~W. Shor.
\newblock Algorithms for quantum computation: Discrete logarithms and
  factoring.
\newblock In {\em IEEE Symposium on Foundations of Computer Science}, pages
  124--134, 1994.

\bibitem{Grover1997}
L.~K. Grover.
\newblock Quantum mechanics helps in searching for a needle in a haystack.
\newblock {\em Physical Review Letters}, 79(2):325--328, July 1997.

\bibitem{Barnum2000}
Howard Barnum, Herbert~J. Bernstein, and Lee Spector.
\newblock Quantum circuits for {OR} and {AND} of {OR}s.
\newblock {\em Journal of Physics A: Mathematical and General},
  33(45):8047--8057, 2000.

\bibitem{Cirac1995}
J.~I. Cirac and P.~Zoller.
\newblock Quantum computations with cold trapped ions.
\newblock {\em Physical Review Letters}, 74(20):4091--4094, May 1995.

\bibitem{Gershenfeld1997}
Neil~A. Gershenfeld and Isaac~L. Chuang.
\newblock Bulk spin-resonance quantum computation.
\newblock {\em Science}, 275(5298):350--356, January 1997.

\bibitem{Blythe2006}
P.~J. Blythe and B.~T.~H. Varcoe.
\newblock A cavity-{QED} scheme for cluster-state quantum computing using
  crossed atomic beams.
\newblock {\em New J. Phys.}, 8(10):231, October 2006.

\bibitem{book:Gallagher2005}
T.~F. Gallagher.
\newblock {\em Rydberg Atoms (Cambridge Monographs on Atomic, Molecular and
  Chemical Physics)}.
\newblock Cambridge University Press, August 2005.

\bibitem{Walther2006}
H.~Walther, B.~T.~H. Varcoe, B.~G. Englert, and T.~Becker.
\newblock Cavity quantum electrodynamics.
\newblock {\em Reports on Progress in Physics}, 69(5):1325--1382, 2006.

\bibitem{Varcoe00}
B.~T.~H. Varcoe, S.~Brattke, M.~Weidinger, and H.~Walther.
\newblock Preparing pure photon number states of the radiation field.
\newblock {\em Nature (London)}, 403(6771):743--746, 2000.

\bibitem{Englert1998}
B.~G. Englert, M.~L\"offler, O.~Benson, B.~Varcoe, M.~Weidinger, and
  H.~Walther.
\newblock Entangled atoms in micromaser physics.
\newblock {\em Fortschritte der Physik}, 46(6-8):897--926, 1998.

\bibitem{Jones2009}
M.~L. Jones, G.~J. Wilkes, and B.~T.~H. Varcoe.
\newblock Single microwave photon detection in the micromaser.
\newblock {\em Journal of Physics B: Atomic, Molecular and Optical Physics},
  42(14):145501, 2009.

\bibitem{Harvey2001}
I.~Harvey.
\newblock Artificial evolution: A continuing saga.
\newblock {\em Lecture Notes in Computer Science}, 2217:94--109, 2001.

\bibitem{Tada2002}
M.~Tada, Y.~Kishimoto, M.~Shibata, K.~Kominato, S.~Yamada, T.~Haseyama,
  I.~Ogawa, H.~Funahashi, K.~Yamamoto, and S.~Matsuki.
\newblock Manipulating ionization path in a stark map: Stringent schemes for
  the selective field ionization in highly excited rb rydberg.
\newblock {\em Physics Letters A}, 303(4):285--291, October 2002.

\end{thebibliography}
\end{document}